\documentclass[article,aps,prd,11pt,notitlepage]{revtex4-1}
\usepackage{amsmath,graphicx,float}

\usepackage{bm}
\usepackage{amssymb}
\usepackage{graphicx,xcolor}
\begin{document}
\def\tr{\rm{Tr}}
\def\la{{\langle}}
\def\ra{{\rangle}}
\def\a{{\alpha}}
\def\e{\epsilon}
\def\q{\quad}
\def\w{\tilde{W}}
\def\t{\tilde{t}}
\def\a{\hat{A}}
\def\h{\hat{H}}
\def\E{\epsilon}
\def\E{\mathcal{E}}
\def\p{\hat{P}}
\def\u{\hat{U}}
\def\n{\\ \nonumber}
\def\j{\hat{j}}
\def\g{\hat{G}}
\def\vc{\underline{c}}
\def\vf{\underline{f}}

\title{Klein paradox for bosons, wave packets and  negative tunnelling times}
\author {X. Guti\'errez de la Cal$^{a}$}
\author {M. Alkhateeb$^{b}$}
\author {M. Pons$^{c}$}
\author {A. Matzkin$^{b}$}
\author {D. Sokolovski$^{a,d*}$} 
\affiliation{$^a$ Departamento de Qu\'imica-F\'isica, Universidad del Pa\' is Vasco, UPV/EHU, Leioa, Spain}
\affiliation{$^b$ Laboratoire de Physique Th\'eorique et Mod\'elisation, CNRS Unit\'e 8089, CY Cergy Paris Universit\'e,
95302 Cergy-Pontoise cedex, France}
\affiliation{$^c$Departamento de F\' isica Aplicada I, Universidad del Pa\' is Vasco, UPV-EHU, Bilbao, Spain}
\affiliation{$^d$ IKERBASQUE, Basque Foundation for Science, E-48011 Bilbao, Spain}
\begin{abstract}
\noindent
\textbf{We analyse a little known aspect of the Klein paradox. A Klein-Gordon boson appears to 
be able to cross a supercritical rectangular barrier without being reflected, while spending there a negative amount of time.
The transmission mechanism is demonstrably acausal, yet  an attempt to construct the corresponding causal
 solution of the Klein-Gordon equation fails.
We relate the causal solution to a divergent multiple-reflections series, and show  that the problem 
is remedied for a smooth barrier, where pair production at the energy equal to a half of the barrier's 
height is enhanced yet  remains finite.}
 \date\today
\end{abstract}
\maketitle
$^*$dgsokol15@gmail.com

The Klein paradox \cite{Kl1}, associated with motion of relativistic particles in a potential high enough to bridge 
the particle-anti-particle gap, has frequently been studied within the original one-particle picture, as well 
as by the methods of quantum field theory (for a comprehensive review see \cite{Kl2} and Refs. therein).
Our present interest in Klein tunnelling stems largely from its relation to the so-called "tunnelling time problem"
{\color{black} (various aspects of the problem were discussed, e.g., in \cite{Muga}, for recent experiments 
see  \cite{Nat1} and  \cite{Nat2})}.
It was known since early 1930's \cite{MacColl} that in non-relativistic quantum mechanics 
the centre of mass (COM) of a wave packet (WP), transmitted across a rectangular barrier by tunnelling
(Fig.1a, dashed)
is advanced relative to the freely propagating one by roughly the barrier's width (see Fig.1c). 
It appears, therefore, that the tunnelled particle is capable of crossing the barrier region almost 
infinitely fast. This would, however, contradict Einstein's relativity, and several authors 
(see, e.g., \cite{DLow}) inquired whether the effect would persist if the Schr\"odinger particle 
were to be replaced by a relativistic one, such as a Klein-Gordon  (KG) boson \cite{KG}.
The effect persists as it should \cite{DLow, DS1}, since its explanation lies elsewhere. Tunnelling transmission 
results from destructive  interference between various delays incurred by the barrier,  and an attempt to determine the delay actually experienced by a particle meets with the same difficulty as the quest to pinpoint the actual slit chosen by an electron in Young's double slit experiment.
Moreover, the displacement  by  the barrier width  $d$ shown in Fig.1c can be related to the so-called "weak values", 
obtained in highly inaccurate quantum measurements, specifically designed to perturb quantum interference
only slightly, and as such cannot be used to deduce the duration a particle spends in the barrier \cite{DS2,DS3}.
\newline
The situation changes for a 
supercritical barrier, i.e., for one whose height exceeds the particle-antiparticle (P-AP) gap (see Fig.1b).
As is well known, spin-zero particles are described in relativistic quantum theory by the Klein-Gordon equation (KGE) (see the textbooks \cite{KG}, \cite{ST} and Refs. \cite{Matz1}  and \cite {ST1} for very recent work.)
\begin{figure}[h]
\includegraphics[angle=0,width=17cm, height= 6cm]{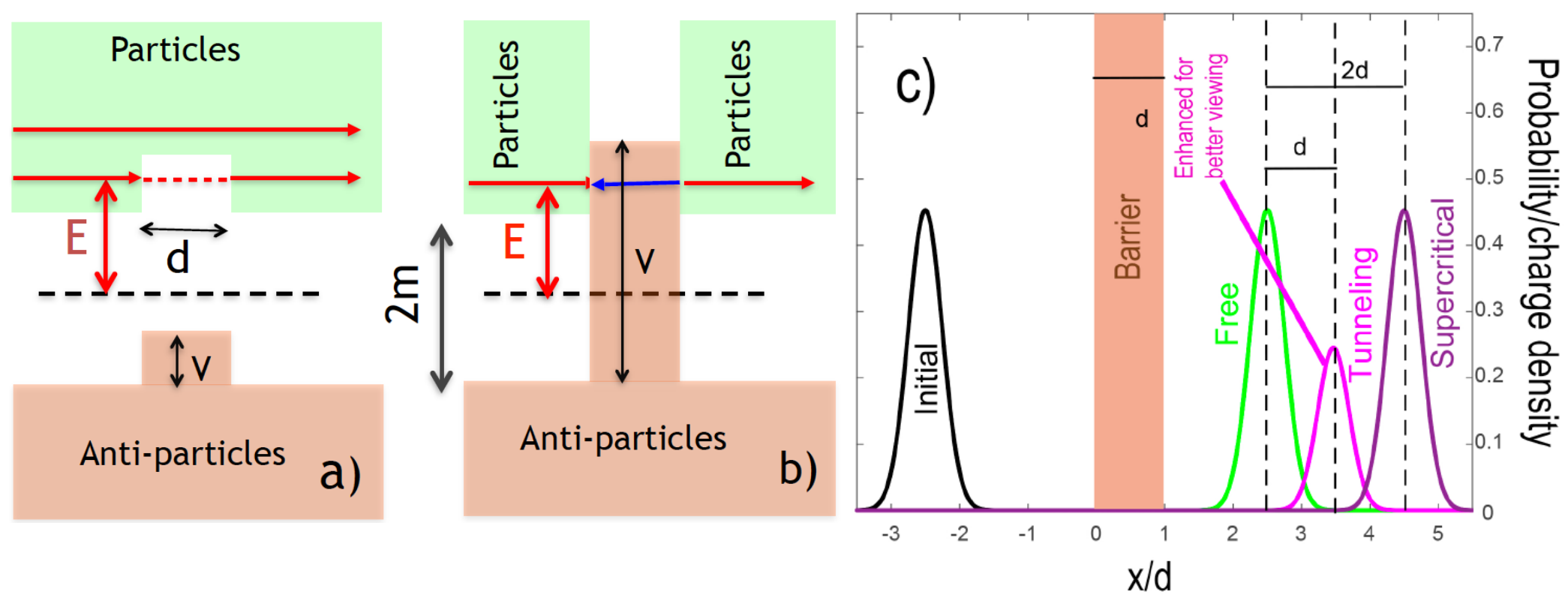}
\caption {Transmission of a Klein-Gordon particle  a) across a rectangular subcritical barrier and 
b) across a supercritical barrier c) Initial and final densities for a free, sub-critical (significanty enhanced) and supercritical 
transmission. All velocities inside and outside the barrier are small compared to the speed of light. (For more details see
Methods).}
\label{fig:FIG0kg}
\end{figure}
Constructing a broad wave packet from the scattering solution of the Klein-Gordon equation (KGE), 
and choosing the mean energy to be a half of the barrier's height, $E=V/2$, 
we find the particle not only transmitted without reflection, 
but also advanced relative to free propagation by {\it twice} the barrier's width $d$ (see Fig.1c).
With motion classically allowed both  outside (for the particles)  and inside the barrier (for the anti-particles),
the trajectory followed by the WP needs to spend inside the barrier a {\it negative} duration 
\begin{eqnarray}\label{Z}
\tau \approx -md/p_0.
\end{eqnarray}
{\color{black} The condition $E=V/2$ defines the so-called super-Klein-tunnelling regime, and we refer the reader to Refs.\cite{SK1}
and \cite{SK2} for further discussion.}
\newline
Following Feynman \cite{FeynQED}, one can see super-Klein tunnelling as a process where
 a particle, scattered back into the past at the left edge of a supercritical barrier, 
 emerges from the barrier's right edge at an earlier moment,  
thus gaining a distance $2d$ on the freely propagating  one.
Alternatively,  one can say that a P-AP  pair is created at the right edge of the barrier {\it before}
the particle reaches it, and the AP travelling to the left  annihilates the original particle,   
which is replaced  by an identical copy already a distance $d$ from the barrier's right edge. 
Both descriptions would benefit from a further discussion.
The idea of travelling back in time is at least strange, 
while the ability of the pair creation mechanism to anticipate the arrival of the incoming particle 
points towards the {\it acausal} nature of the proposed mechanism.
\newline
Furthermore, if the  barrier is extended to the right to form a potential step, any incident particle WP, built with the help of the scattering states, will have to be accompanied by its AP mirror image, moving towards it. 
For a potential step this can be changed by using the scattering states with the sign of AP's momentum reversed \cite{EXPL}, 
so that the incoming particle creates, in an explicitly causal manner, P-AP pairs as it collides with the barrier. 
However, for a barrier of finite width we do not have the freedom 
to reverse the AP's momentum at will, since the solution is already fixed by the
requirement
that only outgoing particles may exist beyond the barrier's right edge.
 It appears that the situation shown in Fig.1c 
is the only one possible for a WP, built from the scattering states continuous with the first derivative 
at the discontinuities of the potential. 
\newline
This, however, leads to another problem. We could equally have decided to solve the KGE by a finite difference method, 
and prepare an incident WP close enough to the barrier for the AP to be already present
if the scattering states expansion were used.
However, in our present construction the AP component is absent, and we must be dealing with a solution different from the one shown in Fig.1c. Assuming the barrier to be wide enough, we can now expect the incident WP first partly penetrate the barrier, and partly be scattered back, as in the case of a potential step. The penetrated part will have to continue until 
scattered off the far edge of the barrier, and so on.  We may call the solution obtained in this manner {\it causal}.
Thus, acting in two different yet perfectly legitimate  ways, 
we arrive at two different solutions to the same problem. 
It remains to see which of these solutions can be considered physical and to establish, if possible, 
a connection between them. We will do it in the rest of the paper.
\section{Multiple reflections in non-relativistic scattering}
It is instructive to consider first  over-barrier transmission of a non-relativistic particle with $E\approx \E(p)+mc^2$, 
$\E(p)\equiv p^2/2m$, $\E,V << mc^2$
 (Fig.1a, solid). 
A solution, describing a particle with 
a mean momentum $p_0$ can be constructed from the scattering solutions $\phi(x,p)$ of the Schr\"odinger equation in the usual way (we put to unity $\hbar=1$ and the speed of light, $c=1$), 
\begin{eqnarray}\label{Z1}
\psi(x,t) = \int dp A(p-p_0)\phi(x,p)\exp(-i\E t).
\end{eqnarray}
For a rectangular potential occupying the region $0\le x\le d$ these states are given by
\begin{eqnarray}\label{Z2}
\phi(x,p) = T(p,q)\exp(ipx),\q x>d\n
               = [B_+(p,q)\exp(iqx) +B_-(p,q)\exp(-iqx)]T(p,q)
               \q 0 \le x \le d \n
               =\exp(ipx)+C(p,q)T(p,q) \exp(-ipx)\q x<0,
\end{eqnarray}
where $T(p,q)$ is the transmission amplitude, $B_{\pm} =T\exp[i(p\mp q)d](1\pm p/q)/2$, 
$C=(1-q/p)B_+/2+(1+q/p)B_-/2$, and 
\begin{eqnarray}\label{Z3}
q(p)=\sqrt {p^2-2mV}
\end{eqnarray}
is the (real) particle's momentum in the barrier region.
In Eq.(\ref{Z1}) $A(p-p_0)$ is chosen so that at some initial time $t=t_i$ $\psi(x,t)$,
built up from the plane waves with positive momenta, lies sufficiently far to the left of the barrier (see  in Fig.1c).
With the help of Eq. (\ref{Z1}) we can monitor the particle's evolution throughout the scattering process 
both outside and inside the barrier.  
\newline
The transmission amplitude of a rectangular potential, a barrier or a well, is well known. Since $|p-q|< p+q$, it can be represented by 
a converging geometric progression, known as the multiple reflections expansion (MRE),
\begin{eqnarray}\label{Z4}
T(p,q)=\frac{4pq\exp(-ipd)}{(p+q)^2\exp(-iqd)-(p-q)^2\exp(iqd)}=\
\q\n
\frac{4pq\exp(-ipd)}{(p+q)^2}\sum_{n=0}^\infty \frac{(p-q)^{2n}}{(p+q)^{2n}}\exp[i(2n+1)qd] 
\equiv \sum_{n=0}^\infty T_n(p,q).
\end{eqnarray}
Inserting the MRE in Eq.(\ref{Z2}) we have $\phi(x,p)=\sum_{n=0}^\infty \phi_n(x,p)$, while for our WP in Eq.(\ref{Z1}) we find
\begin{eqnarray}\label{Z5}
\psi(x,t) = \sum_{n=0}^\infty\psi_n(x,t)=
\sum_{n=0}^\infty   \int dp A(p-p_0)\phi_n(x,p)\exp(-i\E t).
\end{eqnarray}
The usefulness of splitting the scattering solution into sub-amplitudes  $\psi_n(x,t)$ 
is best illustrated by choosing a broad barrier, and a momentum distribution $A(p-p_0)$ sharply 
peaked around $p=p_0$, so that all factors in the MRE (\ref{Z4}), except the exponentials containing 
$d$, can be evaluated at $p=p_0$ and $q(p)=q(p_0) \equiv q_0$. Approximating in the  exponents
\begin{eqnarray}\label{Z6}
q(p)=q(p_0) +\partial_p q(p_0)(p-p_0)...\approx q(p_0)+p_0/q_0(p-p_0)
\end{eqnarray}
turns the r.h.s of Eq.(\ref{Z3}) into a sum of freely propagating wave packets with different 
{ spatial} shifts, weighted by different individual factors. For example, in the region $x>d$, which contains only transmitted particles, we have wave packets moving away from the barrier,
\begin{eqnarray}\label{Z7}
\psi_I(x>d,t) \approx \sum_{n=0}^\infty X_n(p_0, q_0) \psi_0(x-x_n,t), 
\end{eqnarray}
where 
\begin{eqnarray}\label{Z8}
\psi_0(x,t)\equiv \int dp A(p-p_0)\exp(ipx)\exp(-i\E t) 
\end{eqnarray}
is the free WP in the absence of the barrier potential,
\begin{eqnarray}\label{Z9}
x_n(p_0,q_0)=d[1-(2n+1)p_0/q_0] < 0,
\end{eqnarray}
and 
\begin{eqnarray}\label{Z9a}
X_n(p_0,q_0)=\frac{4p_0q_0}{(p_0+q_0)^2}
\frac{(p_0-q_0)^{2n}}{(p_0+q_0)^{2n}}
\exp[i(2n+1)(q_0-p_0^2/q_0)d].
\end{eqnarray}

Thus, the peak of the $\psi_n(x,t)$ appears at $x=d$ at the same time as would the peak 
of a free WP, staring its motion at $t=t_i$, but placed a distance $x_n$ behind the true $\psi_0(x,t_i)$. 
In other words all terms in Eq.(\ref{Z7}) are {\it delayed}, and in a snapshot taken at a time $t$ 
those already present in the region  $d< x < \infty$ would lag behind the freely propagating 
$\psi_0(x,t)$. Free propagation can be used as a reference also in the reflection region 
$x<0$, where we will have to consider delayed free WPs with different shifts, $\psi_0(-x-y_n,t)$, reflected 
about the origin, $x\to -x$. Extension to the barrier region is also straightforward, and we can  
summarise the situation 
as follows.
\begin{figure}[h]
\includegraphics[angle=0,width=14cm, height= 5cm]{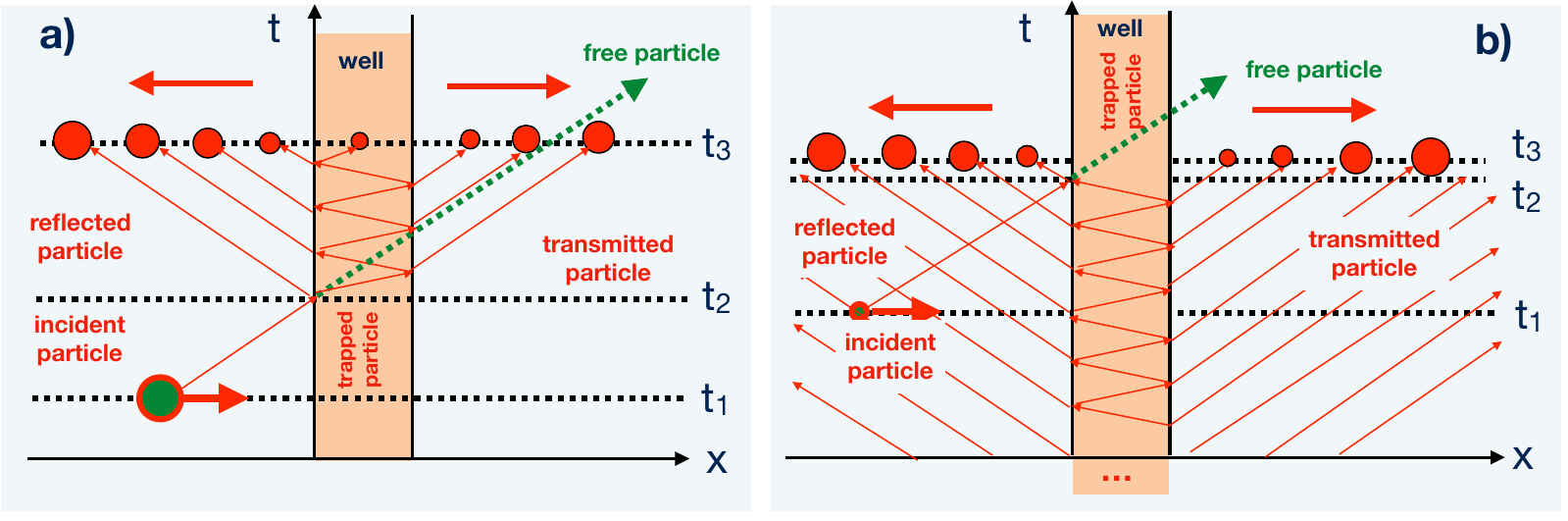}
\caption {Schematic space-time diagram showing the evolution of the bound (a) and unbound (b) solutions of the Schr\"odinger equation for a potential well (\ref{Z9b}). Centres of mass of wave packets are represented by circles, proportional
(although not to scale) to their sizes. The dashed lines map the free motion.}
\label{fig:FIG1}
\end{figure}
The initial wave packet arrives at the left edge of the barrier, and is partly 
reflected, partly transmitted into the barrier's interior, where it continues moving to the right 
until reaching the barrier's right edge. There a part of it escapes, while the rest, reflected back into 
the barrier, moves to the left until impinging on the left edge from inside. 
The cycle repeats itself, 
the probability to remain inside the barrier region steadily decreases, and the scattering 
is finished as $t\to \infty$.
\newline A similar description applies also to a passage over a potential well, 
\begin{eqnarray}\label{Z9b}
W(x)=-V <0, \q \text{for}\q 0\le x \le d 
\end{eqnarray}
with the only difference that now the particle moves faster inside the potential, since $q(p)>p$, 
and some of transmitted WPs  may appear advanced.
{\color{black} This situation is illustrated schematically in Fig.2a, which shows the developments between a time 
$t_1$, when the incident wave packet (green) is approaching the barrier from the left, and a $t_2 >t_1$, 
when the reflected and transmitted wave packets are leaving the barrier region in opposite directions. 
A small part of the wave function continues to move inside the barrier, and the straight lines, branching at the 
barrier edges indicate the paths followed by the wave packets' centres of mass. The size of the circle, representing
a wave packet is chosen proportional (although not to scale) to its norm.}
\newline
This is a solution of the scattering type, since the particle undergoes a transition 
between well defined asymptotically free states. The solution is also normalisable and causal, in the sense 
that all transmitted and reflected wave packets in Fig.2a appear after the arrival of the incident particle. 
\section{Unbound solutions of the Schr\"odinger equation}
A different solution of the Schr\"odinger equation can be obtained as follows. 
Despite containing the square root (\ref{Z3}) the transmission amplitude $T(p,q)$
is single valued in the complex $p$-plane, $T(p,q)=T(p,-q)$ and we are free to change the sign 
of $q$ in the first expression in the r.h.s. of Eq.(\ref{Z4}). However, the same change has 
consequences, if made in the following MRE, as the geometric progression now diverges. 
Next we will  look for an application for this divergent series, obtained by substituting  $q\to -q$,
\begin{eqnarray}\label{Y1}
T(p,q)=
T(p,-q) \leftrightarrow 
\frac{-4pq\exp(-ipd)}{(p-q)^2}
\sum_{n=0}^\infty \frac{(p+q)^{2n}}{(p-q)^{2n}}\exp[-i(2n+1)qd] 
\equiv \sum_{n=0}^\infty T_{n}(p,-q),\q
\end{eqnarray}
by proceeding as in the previous section. For a broad barrier, and a narrow momentum distribution 
$A(p-p_0)$ the new solution takes the form
\begin{eqnarray}\label{Y2}
\psi_{II}(x>d,t) \approx \sum_{n=0}^\infty X_n(p_0, -q_0) \psi_0(x-x_n(p_0,-q_0),t), 
\end{eqnarray}
where $x_n(p_0,-q_0)=d[1+(2n+1)p_0/q_0]>0$ and  $|X_n(p_0, -q_0)|\to \infty$ as $n\to \infty$.
With $\psi_0(x-x_n(p_0,-q_0),t)$ all {\it advanced} 
relative to the free propagation, an infinite number of wave packets populate the $x<0$ and $x>d$ regions 
even before the incident particle reaches the barrier. 
The wave packets  further away from the barrier have larger weights, and the norm of the state diverges.
\newline
The exodus of the wave packets in both directions ends with the arrival of the incident WP 
at the left edge of the barrier. The last WP, which has left the barrier in the positive direction, 
is advanced by $x_0(p_0, -q_0)=d+p_0d/ q_0$, and must have left the barrier $md/q_0$ seconds {\it before}
the incident WP reached $x=0$. Similarly, it can be shown that the last WP moving to the left 
emerged at $x=0$ at about the same time the incident WP arrived there. Now we can reconstruct the 
scenario described by the divergent MRE (\ref{Y1}). 
Initially, there
 is not only a WP approaching the barrier from the left, but also 
another wave packet, already trapped in the barrier region.
  On its own, the second component of the wave function
would produce ever decreasing outgoing WPs as bounces off a potential drop. 
The incoming wave packet is tuned to arrive at the left barrier's edge just as the trapped WP approaches it 
from the other side. The incoming WP is reflected and, in  a coherent manner, takes with it the still trapped part, 
leaving the barrier region empty. [A stationary solution corresponding to this process would have a form 
$\exp(ipx)+r\exp(-ipx)$ for $x<0$ and $t\exp(-iqx)$ for $x>0$.]
{\color{black} The situation is sketched in Fig.2b, where there are now infinitely many wave packets moving away from the potential
in both directions, and the incoming particle terminates, rather then starts, the oscillations in the well.}
\newline
There are  several points of interest.
Firstly, the norm of the solution diverges because the trapped part of the wave function 
was assumed to exist already for $t\to -\infty$, and yet must not vanish when the initial WP arrives. Assuming instead that it was injected into the barrier 
region at some finite $t_i$, would give us a suitable physical solution for all $t>t_i$.
(A similar situation occurs in transmission across a supercritical barrier, and we will discuss it shortly).
Secondly it is clear that the solution in Fig.2b cannot be obtained by using the scattering
states in a conventional way, e.g., by using Eq.(\ref{Z1}). The standard scattering ($S$-matrix) theory 
describes transitions between asymptotically free states, whereas the initial condition for the process
shown in Fig.2b requires, from the very beginning, the presence of the particle oscillating in the barrier region 
(note that there are no bound states in the chosen energy range).
\section{Anti-particle  oscillations in a supercritical barrier}
Two different solutions describing supercritical transmission can be obtained in much the same manner.
The main change is that now for the  AP's momentum in the barrier
we have 
$q(p)=\sqrt{(\epsilon -V)^2-m^2},$
and in the expansion around the particle's mean momentum $p_0$ [cf. Eq.(\ref{Z6})]  the second term is negative, 
\begin{eqnarray}\label{X1}
q(p)=q(p_0) +\partial_p q(p_0)(p-p_0)...\approx q(p_0)-p_0/q_0(p-p_0),
\end{eqnarray}
so that an AP with a momentum $q>0$ moves in the barrier from right to left \cite{KG}.
As in the non-relativistic (subcritical) case we can construct two solutions $\Psi_{I}(x,t)$ and $\Psi_{II}(x,t)$, corresponding to the 
converging and the diverging MRE, respectively (see Figs.3a and 3b). However, previously the causal solution was also the bound one. Here, as we will show, the causal state in Fig.3a  becomes unbound as $t \to \infty$, 
while the bound solution in Fig.(3b) is the one responsible for the acausal advancement by $2d$ shown in Fig.1c.
\begin{figure}[h]
\includegraphics[angle=0,width=17cm, height= 4.2cm]{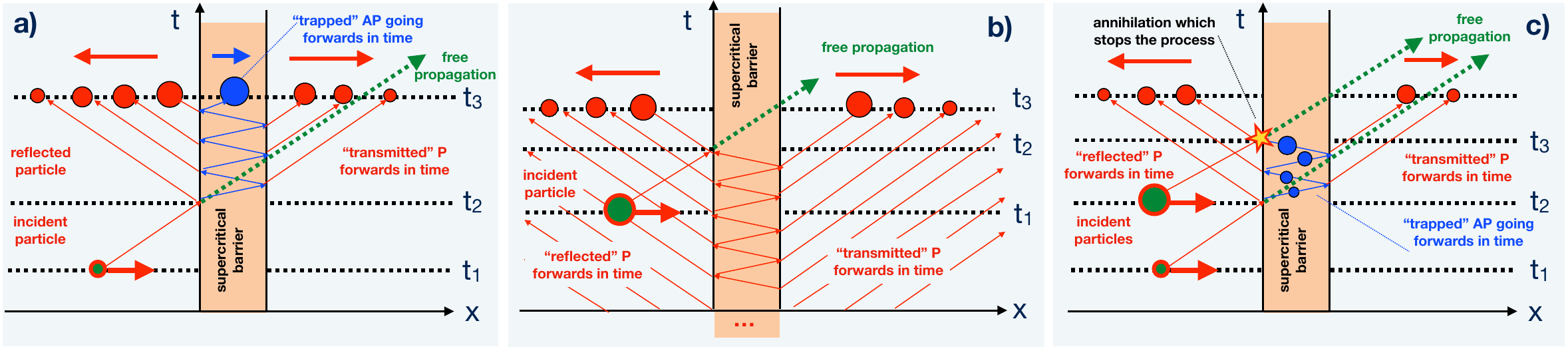}
\caption {a) Same as Fig.2, but for a supercritical barrier. a) Unbound causal solution: AP oscillations are started by the incoming particle.
b) Bound acausal solution: pre-existing AP oscillations are terminated by the incoming particle.
c) A physical solution:  AP oscillations are stated and terminated by two sets of incoming particles.
In all cases we assume $E< V/2$.}
\end{figure}
\newline
 In particular, using the converging MRE series (\ref{Z4}), to the right of the barrier we have
\begin{eqnarray}\label{X2}
\Psi_{I}(x>d,t) \approx \sum_{n=0}^\infty X_n(p_0, q_0) \psi_0(x-x_n(p_0,-q_0),t),
\end{eqnarray}
which differs from Eq.(\ref{Z7}) by replacement $x_n(p_0,q_0) \to x_n(p_0,-q_0)$.
Another novelty is that while previously we had $q< p$, now 
 the velocity of the AP inside the barrier can also be larger or equal to that of the incident particle.
  For $E<V/2$ we have $q>p$, and 
 at $t=t_3$ the $x>d$ region in Fig.3b is populated by an infinite number of advanced  WPs, whose size 
(now representing the mean amount of charge released \cite{EXPL})
{\it decreases}
with the distance from the barrier.
The scenario shown in Fig.3b corresponds to a particle which, having entered the supercritical region at $t=t_2$, begins 
its motion {\it backwards} in time, with ever smaller part of its WP escaping back to normal (forwards in time) propagation after each collision with one of the barrier's edges. In the special case $p_0=q_0$, or $E_0=V/2$, all $X_n(p_0,q_0)$ vanish except for $X_0(p_0,q_0)$, 
and, instead of going down the ladder in Fig.3b, the particle escapes completely on colliding with the right edge of the barrier for the first time. Leaving the barrier {\it before} entering  allows it to gain twice the barrier's width on 
a free particle,  moving only forwards in time (cf. Fig.1c). Thus, the solution $\Psi_{I}(x,t)$ is bound, yet demonstrably acausal. 
\newline
Employing the diverging MRE series yields a causal solution shown in Fig.3a. Now at $t=t_3$ in the range  $x>d$ we have
a finite number of WPs, decreasing in size away from  the barrier,
\begin{eqnarray}\label{X3}
\Psi_{II}(x>d,t) \approx \sum_{n=0}^\infty X_n(p_0, -q_0) \psi_0(x-x_n(p_0,q_0),t),
\end{eqnarray}
which appear only after the initial WP has reached the barrier at $t=t_2$. The scenario shown in Fig.3a corresponds 
to incident particle creating P-AP pairs after striking the left edge of the barrier. The AP wave packet begins
its motion forwards in time, creating more pairs and increasing in size  after each reflection off a barrier's edge. 
This solution is explicitly causal, but unbound, as the charge outside the barrier grows exponentially 
and becomes infinite as $t\to \infty$.
The total charge is  conserved, since the charge density of an AP plane wave propagating in the barrier,
$\varphi(x,t)=\exp(\pm iqx -i\epsilon t)$, given by \cite{KG}
 (we use unit particle's charge, $e=1$)
\begin{eqnarray}\label{X3a}
\rho(x)=(2m)^{-1}[ i(\varphi^*\partial_t\varphi - \varphi \partial_t\varphi^*)-2V\varphi^*\varphi] 
\end{eqnarray}
is of the opposite sign to the particle's charge $e$.
One can, therefore, expect that, with enough charge accumulated inside, the barrier 
will be lowered, and supercriticality will come to an end. This outcome cannot, however, be described by our linear  single-particle model, and is beyond the scope of this paper.
\newline 
We can also give an alternative description of the evolution shown in Fig.3b, similar to the explanation given to 
the evolution in Fig.2b. 
A supercritical potential is unstable with respect to introduction of antiparticles with $E>m$ into the barrier region,  since this leads to pair production.
In Fig.3a exponentially growing 
 AP oscillations, accompanied by emission of particles, are set off by the arrival of the incident particle.
In Fig.3b such oscillations, present from the very beginning, (we recall that a P going backwards in time is equivalent to an AP going forwards) are terminated when a particle wave packet arrives just in time 
to annihilate the antiparticle content.
Solutions $\Psi_{I}$ and $\Psi_{II}$, well suited to describe the onset and termination of the AP oscillations
in the barrier, can be combined into a  more "physical" solution sketched in Fig.3c.
There the barrier oscillations, started by the initial particle, are later terminated, annihilated by yet more incoming particles. This would require an initial state consisting of two well separated wave packets \cite{CAT}, 
which would first start and then later end the process. 
\section{The singularity at $E=V/2$}
For a rectangular barrier, construction of the causal solution meets with an additional difficulty.
While in the special case $p_0=q_0$ [$E(p_0)=V/2$]  the $\psi_I$ corresponds to perfect transmission with an advancement 
shown in Fig.1c, the causal solution $\psi_{II}$ in Eq.(\ref{X3}) becomes infinite, as the integrals over $p$ [cf. Eq.(\ref{Z5})]
now diverge due to the presence of a pole at $p=q$ in Eq.(\ref{Y1}). 
This is caused by the sharp drops of a rectangular potential at $x=0$ and $x=d$, and can be remedied, 
e.g., by using instead a combination of two hyperbolic tangent potential  steps, 
\begin{eqnarray}\label{V2}
W_{smooth}(x) = V[ \tanh(bx) - \tanh[b(x-d)]/2,
\end{eqnarray}
for which the transmission amplitude  was given in \cite{TANH}.
Now the potential varies over a region $\delta x \sim 1/b$, and the Klein tunnelling persists for as long as
$\delta x$ is small compared to the particle's Compton wavelength, $\delta x < \hbar /mc$.  
Using the results of  \cite{TANH} it can be shown (see Methods) that for a smooth potential (\ref{V2}),
with  $\delta x << d$, one can continue using the MRE (\ref{Y1}), with 
the pole in $T_n(p,-q)$ moved into the complex $p$-plane, 
\begin{eqnarray}\label{V3}
(p-q)^{-1}\to (p-q+i\delta)^{-1}, \q \delta=V^2/2b.
\end{eqnarray}
Thus, the pair production \cite{EXPL} at $p=q$ ($E=V/2$) is enhanced, but remains finite.
The causal solution 
 for a WP, whose width in the momentum space is small compared to $\delta$, 
is still given by Eq.(\ref{X3}), with now finite coefficients $X_n(p_0, -q_0)$.
Figures 4a and 4b show the acausal and causal wave packet solutions for a smooth supercritical  potential (\ref{V2})
at $p_0=q_0$. Figure 5 shows a "physical" situation, in which anti-particle oscillations in the barrier, initiated by 
 the first particle, are quenched after one cycle by arrival of the second wave packet.
 These three solutions of the Klein-Gordon equations are obtained by integration in the momentum space, as in Eq.(\ref{Z5}). (See Methods for more details).
\begin{figure}[h]
\includegraphics[angle=0,width=17cm, height=8cm]{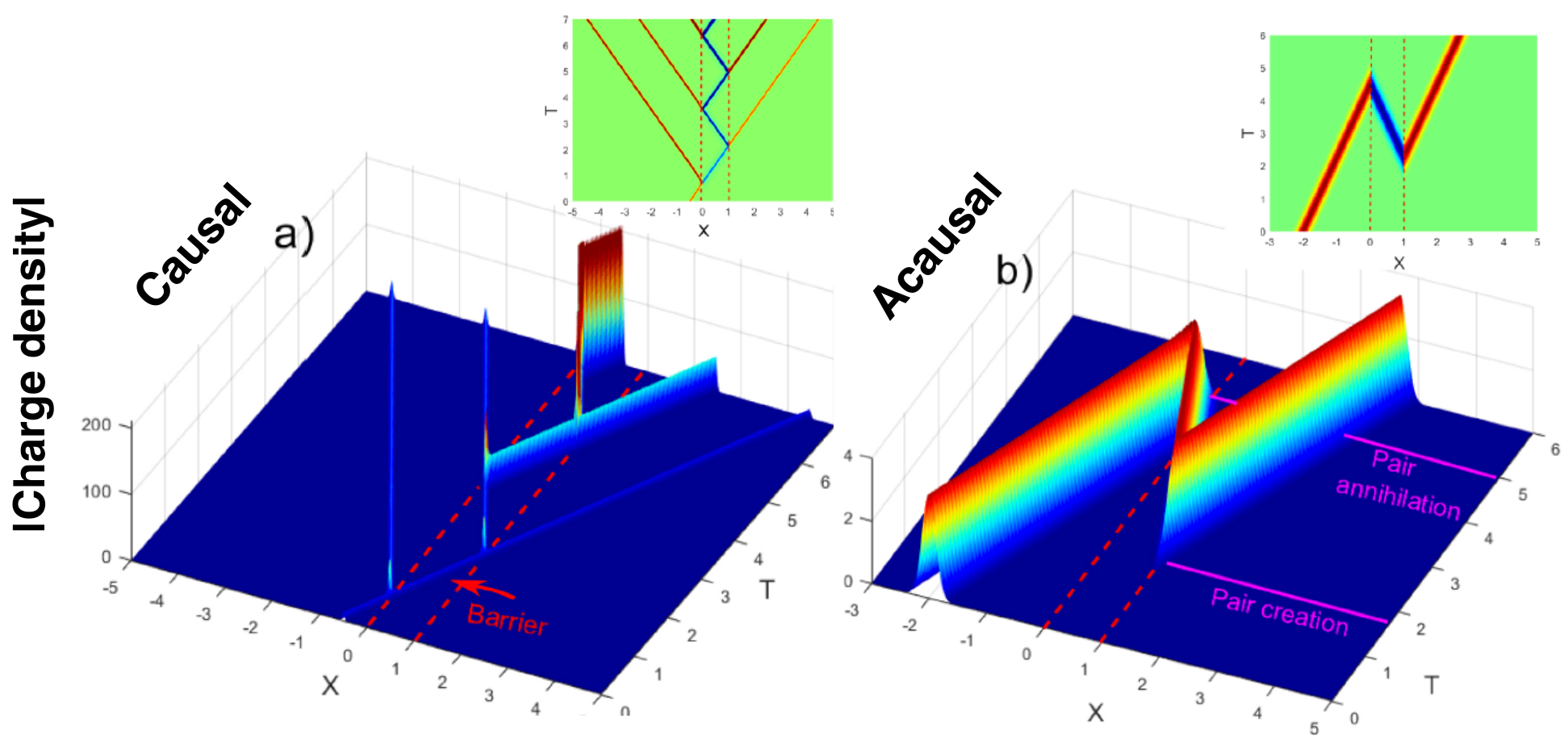}
\caption {
a) Causal evolution (see Fig.3a) of the absolute value of the charge density (\ref{X3a}) for a smooth supercritical potential (\ref{V2}) and $E(p_0)=V/2$  (vs. $X=x/d$ and $T=tc/d$). Anti-particle oscillations in the barrier are started 
by the arrival of the initial wave packet. The charge density, positive outside and negative inside the barrier is shown in the inset. 
b) Same as a), but for the acausal evolution (see. Fig.3b), resulting in the advancement by $2d$ shown 
in Fig.1c.  For the details of the calculations see Methods. }
\end{figure}
\begin{figure}[h]
\includegraphics[angle=0,width=12cm, height=10cm]{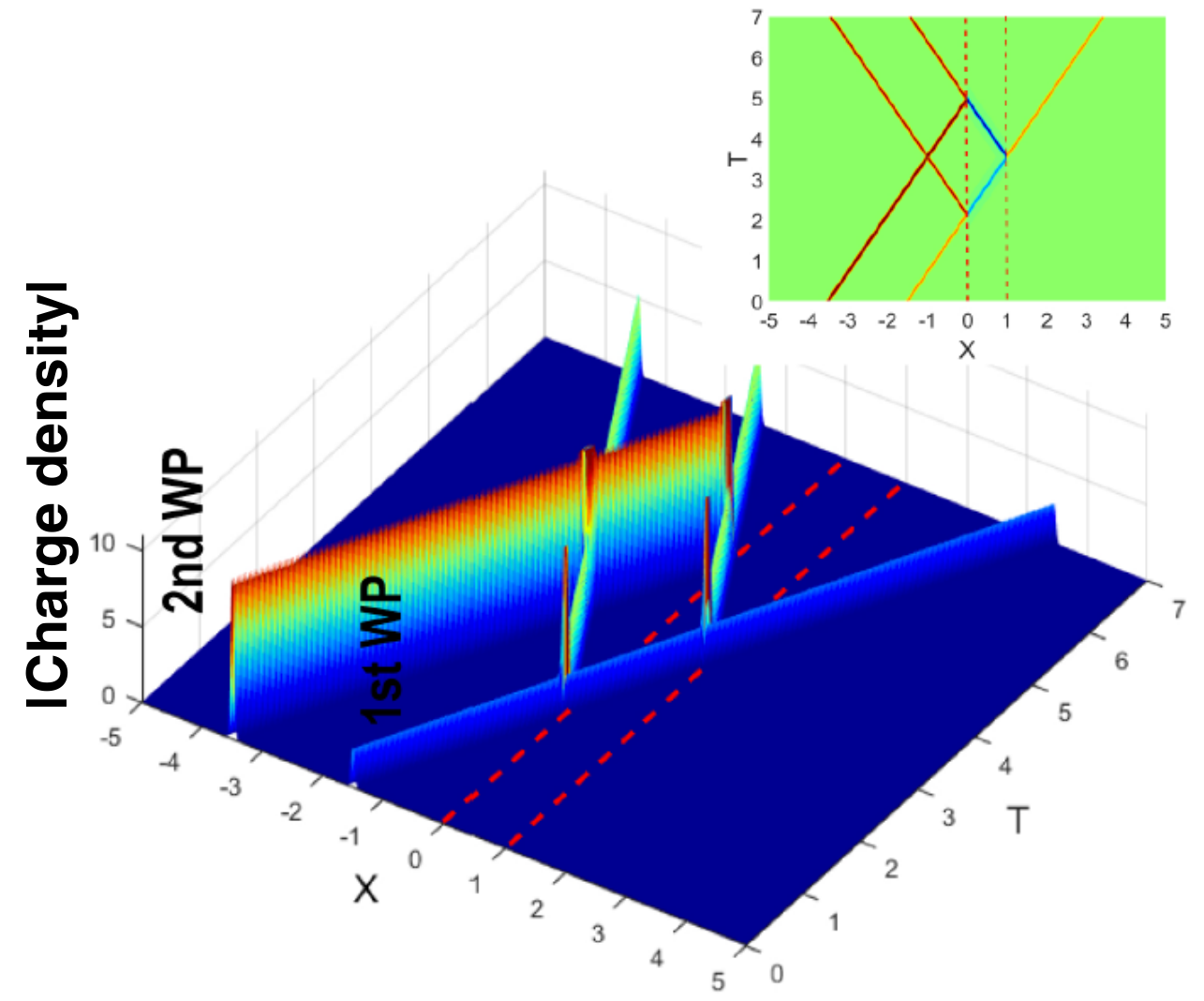}
\caption {A two-wave packet solution (see Fig. 3c) for a smooth supercritical barrier (\ref{V2}) and $E(p_0)=V/2$.
Anti-particle oscillations, started by the arrival of the first wave packet (WP) are quenched after one full cycle
by the arrival of the second WP. The charge density, positive outside and negative inside the barrier is shown in the inset. For the details of the calculations see Methods.}
\end{figure}
\section{Summary and discussion}
To understand the origin of the negative tunnelling time implied by Fig.1c it is useful to consider first 
the passage of a  non-relativistic particle over a broad rectangular potential, a barrier or a well.
A scattering solution for a potential with such sharp edges is determined
by the transmission and reflection 
amplitudes, and two coefficients multiplying the plane waves in the potential region. All four amplitudes can be 
expanded according to the number of reflections experienced by the particle as it crosses the potential.
Two such expansions can be constructed in the form of a convergent and divergent geometric progressions. 
A wave packet introduced into the barrier region performs decaying oscillations between the barrier edges,
which end when the particle finally escapes. The converging expansion describes such oscillations, initiated 
by an initially free incoming particle, which end with a free escaped particle travelling in either direction as $t\to \infty$. 
The corresponding solution, apparently causal and bound, can be obtained by means of an $S$-matrix theory, 
e.g., by evaluating the momentum space integrals (\ref{Z1}).  
\newline
Perhaps unexpectedly, the diverging series also has an  application. It describes a somewhat bizarre situation of a particle in a state, initially containing
 (at $t_1 > -\infty$)
  not only  
a free WP approaching the potential, but also another WP 
already oscillating
 in the potential region. The incident WP cancels the oscillations by taking with it, while being reflected,
 what amplitude remained in the barrier or the well region. Since the particle is not free asymptotically, this is not a conventional scattering situation, the state becomes unbound if $t_1\to -\infty$, and may be described as "acausal". 
This second scenario is difficult to realise in practice, and is usually ignored in non-relativistic quantum mechanics. 
Still it is a possible one, at least in principle, and quantum theory has a way of describing it. 
\newline
Such a scenario, however, is not easily dismissed for a  scalar relativistic particle incident on a supercritical 
barrier, in whose interior the anti-particles are allowed to move. In this case, the bound scattering solution turns out to be also the acausal one.  The causal solution, on the other hand, becomes unbound as $t \to \infty$. This is because an antiparticle WP 
introduced into the barrier region must perform exponentially growing oscillations, with each collision 
with a  barrier's edge producing pairs of outgoing particles, and anti-particles, reflected back into the barrier. 
Such particle-antiparticle oscillations can be started by an incident particle arriving at the barrier from outside, and subsequently reflected, as shown in Figs.3a and 4a. One-particle Klein-Gordon theory lacks the means to put an end to the oscillations, except by annihilating the AP content in the barrier by yet more particles arriving at the barrier from outside. Without this, the (negative) charge inside the barrier, and the (positive) 
charge outside it,  grow unchecked and the solution becomes unbound as $t\to \infty$.
This is a scenario obtained by using the {\it divergent} MRE. 
\newline
By using the  {\it convergent} MRE, one obtains a finite yet acausal solution, describing a process in which 
anti-particle oscillations previously existing the barrier region are terminated by the arrival of more particles, 
as shown in Figs.3b and 4b. In this case the solution, which is finite and normalisable, can be described by the standard scattering 
theory, provided the AP is seen as a particle moving back in time and loosing, rather than acquiring, amplitude
after each collision with a barrier's edge.  
\newline
Finally, unlike in the non-relativistic case, the 
momenta outside and inside a supercritical barrier become equal for $E=V/2$, in which case only the direct ($n=0$) term survives in the convergent MRE (\ref{Z4}). This term is responsible for a single-passage acausal evolution in Fig. 4b, and the "negative duration" 
(\ref{Z}), spent by the time-reversed particle's trajectory, which crosses the barrier only once.
 \begin{center}
\textbf{Acknowledgements}
\end{center}
Financial support of
MCIU, through the grant
PGC2018-101355-B-100(MCIU/AEI/FEDER,UE) (XGdC, MP, DS), of Spanish MINECO, project FIS2016-80681-P (MP), and of the Basque Government Grant No IT986-16 (MP, DS)
is gratefully acknowledged. 
\begin{center} 
\textbf{Author contributions}
X.G de la C, M.A, M.P.,A.M.,  and D.S. all  wrote the paper, and reviewed it.
\end{center}
\begin{center} 
\textbf{Competing interests}
The authors declare no competing interests.
\end{center}

 \newpage
 \section{Methods}
 \subsection{The multiple reflections expansions (MREs)}
 Consider a broad potential, a barrier or a well, equal to $V$  for $0<x<d$, and zero otherwise, 
 and smoothly changing between these values in a small vicinity of $x=0$ and $x=d$.
Starting from the right of the barrier, we assume that for $x>d$ there is only an outgoing wave 
$ \exp (ipx)$.
Matching the solutions 
 gives ($q$ is the momentum in the potential region)
$$0<x<d:\q  a_+\exp(iqx)+a_-\exp(-iqx) \gets \exp(ipx).$$
$$ x<0: \q b_{++}\exp(ipx) + b_{-+}\exp(-ipx) \gets \exp(iqx)$$
$$ x<0: \q b_{+-}\exp(ipx) + b_{--}\exp(-ipx) \gets \exp(-iqx),$$
and the solutions on both sides of the barrier are connected, 
$$ x<0: \q (b_{++}a_++b_{+-}a_-)\exp(ipx)+(b_{-+}a_++b_{--}a_-)\exp(-ipx) \leftarrow \exp(ipx) \q :x>d$$
The transmission and reflection amplitudes are, therefore, given by 
\begin{eqnarray}\label{AA1}
T(p,q) = 1/(b_{++}a_++b_{+-}a_-)\q \text{and} \q R(p,q)=\frac{b_{-+}a_++b_{--}a_-}{b_{++}a_++b_{+-}a_-}.
\end{eqnarray}
Similarly, inside the potential region we have
\begin{eqnarray}\label{AA2}
 \psi(x,p) =\frac{a_+\exp(iqx)}{b_{++}a_++b_{+-}a_-} +\frac{a_-\exp(-iqx)}{b_{++}a_++b_{+-}a_-}
\equiv
B_+(p,q)\exp(iqx)+B_-(p,q)\exp(-iqx).\q
\end{eqnarray}
The factor $(b_{++}a_++b_{+-}a_-)^{-1}$, present in all amplitudes $T$, $R$, and $B_{+-}$, 
can formally be represented as geometric progressions,
\begin{eqnarray}\label{AA3}
\frac{1}{b_{++}a_++b_{+-}a_-} \to \frac{1}{b_{++}a_+}\sum_{n=0}\left (\frac{b_{+-}a_-}{b_{++}a_+}\right )^n \text{or} \q
\to \frac{1}{b_{+-}a_-}\sum_{n=0}\left (\frac{b_{++}a_+}{b_{+-}a_-}\right )^n, 
\end{eqnarray}
one of which will converge, and the other diverge. With these we can construct an MRE (\ref{Z5})
for the wave function both inside and outside the potential.
\subsection{A rectangular potential}
In particular, for a rectangular potential $W(x)=V$, $0\le x \le d$, and $0$ otherwise, we have 
\begin{eqnarray}\label{AA4}
a_+=\exp[i(p-q)d](1+p/q)/2,\q a_-=\exp[i(p+q)d](1-p/q)/2\q\q\q\q\n
b_{++}(p,q)=(1+q/p)/2= b_{--}(p,q),\q  b_{-+}(p,q)=(1-q/p)/2=b_{+-}(p,q).
\end{eqnarray}
Divergent MREs for the transmission amplitudes are  then given by  
\begin{eqnarray}\label{AA5}
T(p,-q)\sim  \frac{1}{b_{+-}a_-}\sum_{n=0}\left (\frac{b_{++}a_+}{b_{+-}a_-}\right )^n =
\frac{-4pq\exp(-ipd)}{(p-q)^2}\sum_{n=0}^\infty \frac{(p+q)^{2n}}{(p-q)^{2n}}\exp[-i(2n+1)qd],\q
\end{eqnarray}
and
\begin{eqnarray}\label{AA6}
R(p,-q)\sim\left [\frac{b_{--}(p,q)}{b_{+-}(p,q)} +\frac{a_{+}(p,q)}{a_{-}(p,q)}\right ]
\sum_{n=0}\left (\frac{b_{++}a_+}{b_{+-}a_-}\right )^n\n
=
\frac{p+q}{(p-q)}-\frac{4pq}{(p+q)^2}\sum_{n=1}^\infty \frac{(p+q)^{2n}}{(p-q)^{2n}}\exp(-2inqd),
\end{eqnarray}
where the first term corresponds to the direct reflection off the left edge of the barrier, 
while the sum accounts for multiple crossings of the barrier region.
Similar expansions are easily obtained also for the coefficients $B_+$ and $B_-$.
Convergent MREs are obtained from Eqs.(\ref{AA5}) and (\ref{AA6}) by replacing $q\to -q$.
\subsection{A smooth (hyperbolic tangent) step potential }
To obtain the MREs for a smooth potential 
$ W_{smooth}(x) = V[ \tanh(bx) - \tanh[b(x-d)]$ we use the connection rules for a smooth step
$W(x)=V \tanh(bx)$ in \cite{TANH}, namely
\begin{eqnarray}\label{AA7}
x<<0: A(p,q)\exp(ipx)+B(p,q)\exp(-ipx) \gets \exp(iqx) : x>>0
\end{eqnarray}
where
\begin{eqnarray}\label{AA7}
A(p,q)= \frac{\Gamma(1-iq/b)\Gamma(-i2/b)}{\Gamma(-ip/2b-iq/2b +\lambda)\Gamma(1-ip/2b-iq/2b-\lambda)},\n
B(p,q)=\frac{\Gamma(1-iq/b)\Gamma(ip/b)}{\Gamma(iq/2b-ip/2b +\lambda)\Gamma(1+ip/2b-iq/2b-\lambda)}, 
\end{eqnarray}
where $\Gamma(z)$ is the Gamma function \cite{Abram}, $q$ and $p$ are the momenta at $x\to \pm \infty$, 
and $\lambda \equiv [b+\sqrt{b^2-V^2}]/2b$ (note a error  in the sign of the first $\lambda$ in the 
the denominator of $B(p,q)$ in \cite{TANH}.) From Eqs.(\ref{AA7})
we have
\begin{eqnarray}\label{AA8}
a_+=-A\exp[i(p-q)d] /(|A|^2-|B|^2), \q a_-=B^*\exp[i(p+q)d]/(|A|^2-|B|^2),\n
b_{++}=A, \q b_{-+}=B, \q b_{+-}=B^*, \q b_{--}=A^*,\q\q\q\q
\end{eqnarray}
which reduce to the results for a rectangular barrier in the limit $b\to \infty$.
Inserting these into Eqs.(\ref{AA1})-(\ref{AA2}) allows us to evaluate the wave function everywhere 
except in small vicinities of $x=0$ and $x=d$.
\newline
We note also that the transmission and reflection amplitudes of a smooth step, 
$t(p,q)=1/A(p,q)$ and $r(p,q)=B(p,q)/A(p,q)$, 
after replacing $q\to -q$ develop an additional pole in the complex $p$-plane,  
since $\Gamma(1-ip/b+iq/b-\lambda)\to \infty$ as $1-ip+iq-b\lambda\to 0$. Recalling that
$\Gamma(z) \sim 1/z$, for $b^2 >> V$, we can approximate 
\begin{eqnarray}\label{AA9}
t(p,-q) \approx \frac{2p}{p-q+i\delta}, \q \delta = V^2/2b.
\end{eqnarray}
This pole becomes a problem in the limiting case of a rectangular barrier, since it moves to the real $p$-axis
in the limit $b\to \infty$. Equation (\ref{AA9}) also provides a general rule ($p-q \to p-q+i\delta$) for treating 
poles in Eqs.(\ref{AA5}) and (\ref{AA6}). The causal solution 
at $E(p_0)=V/2$ remains, therefore, finite for a potential with smooth edges. 
\subsection{The solution shown in Fig. 3c }
To construct the solution, we choose a momentum distribution $A_0(p)$, and evaluate the causal solution $\psi_{II}$ up to the moment when, after $n-1$ full oscillations in the barrier the AP wave packet,
$\psi_n(x,t)=\int dp A_n(q)\exp(iqx-iEt)$ is travelling towards the barrier's left edge. The momentum distribution of the WP, $A_n(q)$ is easily found with the help of Eqs.(\ref{AA2}) and (\ref{AA3}).  We need the second incident WP, with   $A_1(p)$ such that, after being reflected $\psi_1=\int dp A_1(p) \exp(ipx-iEt)$ leaves the barrier empty (or, if one prefers, such that the AP oscillations set off by its arrival will cancel the ones that already exist in the barrier).
Using scattering states  $\varphi(p,q)$ for the potential step at $x=0$, 
\begin{eqnarray}\label{AC1}
x<0: =C_1(p,q)\exp(ipx)+C_2(p,q)\exp(-ipx)\gets \exp(iqx) \q : x>0. 
\end{eqnarray} 
we have $A_1(p)=C_1(p,q)A_n(q).$  Now, at $t\to -\infty$, an initial sate  
\begin{eqnarray}\label{AC2}
\psi_{in}(x,t)=\int dp [A_0(p)+A_1(p)]\exp(ipx-iEt)
\end{eqnarray} 
consists of two well separated wave packets, the first of which starts, and the second cancels 
propagation of the anti-particles in the barrier region.
\subsection{Figures 1c, 4, and 5}
In dimensionless variables $X=x/d$, $T=tc/ d, \q W=dV/\hbar c$ and $M=mdc/\hbar$
the Klein-Gordon equation takes the form
$$[(i\partial_T-W)^2+\partial^2_X]\varphi =M^2\varphi,$$
and for the dimensionless charge density ($e=1$) we have 
$$\rho'\equiv (2M)^{-1}[ i(\varphi^*\partial_T\varphi - \varphi \partial_T\varphi^*)-2W\varphi^*\varphi] =
\rho d,$$
with $\rho$ given in Eq.(\ref{X3a}).
A Gaussian momentum distribution $A$ becomes
$A(P-P_0)=const\times \exp[(P-P_0)^2/\Delta P^2 +iPX_0]$
where  $P= pd$, $P_0=p_0 d$, $\Delta P =\Delta p  d$, 
and $X_0=x_0/d$. Finally, the momentum in the barrier is given by
$Q=
\sqrt {(\sqrt{P^2-M^2}-W)^2-M^2}$, while for the energy we find $E(P)=\sqrt{P^2+M^2}$.

Fig.1c was obtained by integration of Eq.(\ref{Z1}) in the momentum space for 
 $M=2\cdot 10^3$ and a rectangular barrier with
$W=0$ (free), $W/M= 5\cdot 10^{-4}<< 1$ (subcritical), and $W/M = 2 $ (supercritical), 
with $P_0=20$, $\Delta P=4$, and $X_0=-5/2$. The final WPs are shown for $T=5$.
\newline
Fig.4a was obtained by integration of Eq.(\ref{Z1}), for a smooth barrier (\ref{V2}), 
with $M=5\cdot 10^3$, $W/M= 2.2361$, $bM/d=100$, for $\Delta P=25/3$, $X_0=-2$ and $\ E(P_0)/W=1/2$.
\newline  
Fig.4b was obtained by integration of Eq.(\ref{Z1}), for a smooth barrier (\ref{V2}), 
with $M=2\cdot 10^4$, $W/M=2\sqrt{2}$, $bM/d=2$, for $\Delta P=100/3$, $X_0=-0.5$ and $ E(P_0)/W=1/2$. 
\newline
Fig.5  was obtained by integration of Eq.(\ref{Z1}), for a smooth barrier (\ref{V2}), 
with $M=2\cdot 10^4$, $W/M=2\sqrt{2}$, $bM/d=2$, for $\Delta P=100/3$, $X_0=-0.5$ and $E(P_0)/W=1/2$. 
Initial momentum distribution corresponds to two WPs, incident on the barrier,
$A_0(p)=A(P-P_0)
+ A_1(P)$ [cf. Eq.(\ref{AC2})]. 
In all cases we used $\hbar=1$, $c=1$ and $e=1$. 
\end{document}